\documentstyle[aps,prl]{revtex}
\begin{document}
\twocolumn[
\title{Phase diagram of the integer quantum Hall effect in p-type Germanium}

\author{M. Hilke, D. Shahar$ ^{1}$, S.H. Song$ ^2$, D.C. Tsui, and
Y.H. Xie$ ^3$$^{,4}$}
\address{Dpt. of Elect. Eng., Princeton University, Princeton, New
Jersey, 08544\\
$ ^{1}$Present address: Dept. of Condensed Matter Physics, Weizmann
Institute,
Rehovot 76100, Israel\\
$ ^2$Present address: University of Seoul, Seoul, 130-743, Korea\\
$ ^3$Bell Laboratories, Lucent Technologies, Murray Hill,  New
Jersey, 07974\\
$ ^4$Present address: Dept. of Mat. Science and Eng.,
UCLA, Los Angeles, California, 90095}

\date{June 14, 1999}
\begin{center}
\parbox{15cm}{\small

We experimentally study the phase diagram of the integer quantized
Hall effect, as a function of density and magnetic field. We used
a two dimensional hole system confined in a Ge/SiGe quantum well,
where all energy levels are resolved, because the Zeeman splitting
is comparable to the cyclotron energy. At low fields and close to
the quantum Hall liquid-to-insulator transition, we observe the
floating up of the lowest energy level, but {\em no floating} of
any higher levels, rather a merging of these levels into the
insulating state. For a given filling factor, only direct
transitions between the insulating phase and higher quantum Hall
liquids are observed as a function of density. Finally, we observe
a peak in the critical resistivity around filling factor one.

 }

\end{center}
\maketitle ]

Two dimensional electronic systems, subject to a perpendicular
magnetic field exhibit a large variety of phenomena. The most
interesting one, is the quantization of the Hall resistivity,
$\rho_{xy}$, in terms of the quantum unit of resistance $h/e^2$.
In a pioneering work, Kivelson, Lee and Zhang (KLZ) \cite{klz},
proposed a theoretical global phase diagram (GPD) of the quantum
Hall effect expressed in terms of ``laws of corresponding
states''. Their phase diagram is composed of two different types
of stable zero temperature (T) phases, the quantum Hall liquid
phases and the insulating  phase. In a quantum Hall liquid phase,
the diagonal resistivity, $\rho_{xx}=0$, and the Hall resistivity
$\rho_{xy}=h/\nu e^2$. (We will only consider the case in which
the filling factor $\nu$ is an integer.) In the insulating phase
$\rho_{xx}$ diverges at zero {\em T}. The different phases in
KLZ's GPD are a function of disorder strength and magnetic field
(B). At high enough disorder only the insulating phase is present
at finite {\em B}. At lower disorder, when different quantum Hall
liquids exist, only transitions between the $\nu=1$ quantum Hall
liquid and the insulating phase are allowed. This means, for
example, that the $\nu=3$ state has to first undergo a transition
to $\nu=2$ and $\nu=1$ before reaching the insulating phase. This
rule relies on the ``floating up'' of the extended states at
vanishing {\em B}, first discussed by Khmel'nitzkii and Laughlin
\cite{khmelinski}. The ``floating'' argument was later questioned
by recent experiments, in which direct transitions between
$\nu\geq 3$ states and the insulating phase have been observed
\cite{krav,song,chang}.

During the past years, there has been a number of theoretical and
experimental papers focusing on the existence of the ``floating
up'' of extended states. Some experiments in n-type GaAs
\cite{glozman} and in Si-MOSFET \cite{shashkin,okamoto} are
consistent with the ``floating up'' scenario and other experiments
in SI-MOSFET \cite{krav} and in p-type GaAs are inconsistent with
it. Theoretically, this issue is also under debate
\cite{fogler,kunyang}. To analyze this aspect further, we present
new results for a different system, i.e., a two dimensional hole
system (2DHS) confined in a Ge/SiGe quantum well \cite{hilke}, in
which the lowest energy level ``floats up'' but {\bf no} higher
energy level.

Addressing this issue experimentally, involves extracting the zero
{\em T} physics. A phase is insulating, when $\rho_{xx}$ diverges
as $T\rightarrow 0$ as opposed to the quantum Hall liquid phases,
where $\rho_{xy}\rightarrow h/\nu e^2$ and $\rho_{xx}\rightarrow
0$ with vanishing $T$. Therefore, the transition can be defined as
the {\em B}-field, where the {\em T}-dependence of $\rho_{xx}$
changes direction \cite{jiang,shahar}. This method was later
successfully used by several groups to determine parts of the
phase diagram \cite{song,chang,hilke,jiang,shahar,shahar2}. We
will follow the same method and discuss the differences compared
to other methods in connection with our data.

In order to illustrate our procedure, we have plotted in fig. 1a,
$\rho_{xx}$ and $\rho_{xy}$ as a function of $B$ for different
{\em T} 's, ranging from 0.4 K to 4.2 K. The insulating-to-quantum
Hall liquid transition at $B_C^L$ can easily be identified from
the crossing point of the $\rho_{xx}$ curves at different {\em T}
's. With increasing $B$, $\rho_{xy}$ first develops a plateau at
$\nu=3$, then at $\nu=2$ and finally at $\nu=1$. This is an
example of a direct transition between the $\nu=3$ state and the
insulating phase at $B<B_C^L$. At higher fields, the quantum Hall
liquid $\nu=1$ state-to-insulator is shown in fig. 1b. Here again
the transition at $B_C$ is obtained from the crossing point of the
$\rho_{xx}$ traces. In both cases (fig. 1a and fig. 1b)
$\rho_{xy}$ shows no features across the transition. In the case
of $B_C$, $\rho_{xy}$ even remains quantized as was noted
previously \cite{hilke3,hilke4}. In the inset of fig. 1b, we
observe that the transition does not occur at the maximum of the
diagonal conductivity, $\sigma_{xx}$. On the other hand, at $B_C$
and at $B_C^L$, the Hall conductivity, $\sigma_{xy}$, is {\em
T}-independent. One could, therefore, alternatively use the {\em
T}-dependence of $\sigma_{xy}$ to define the transition point
\cite{hughes}, but this method has the disadvantage, that
$\sigma_{xy}$ is not measured directly, because the conductivities
are obtained by inverting the resistivity tensor.

\input epsf
\begin{figure}
\epsfysize=7cm \epsfbox{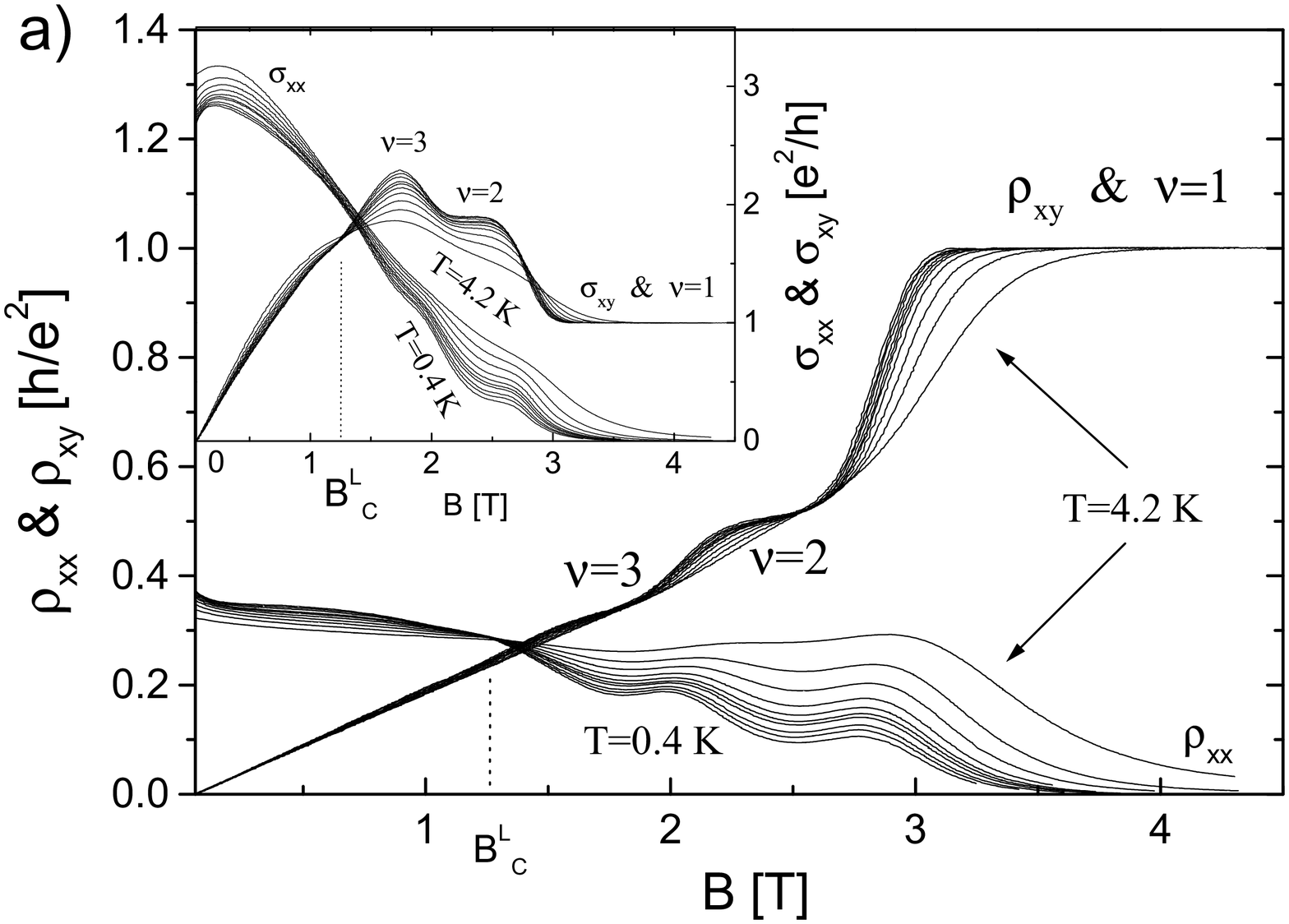} \vspace*{0cm} \epsfysize=7cm
\epsfbox{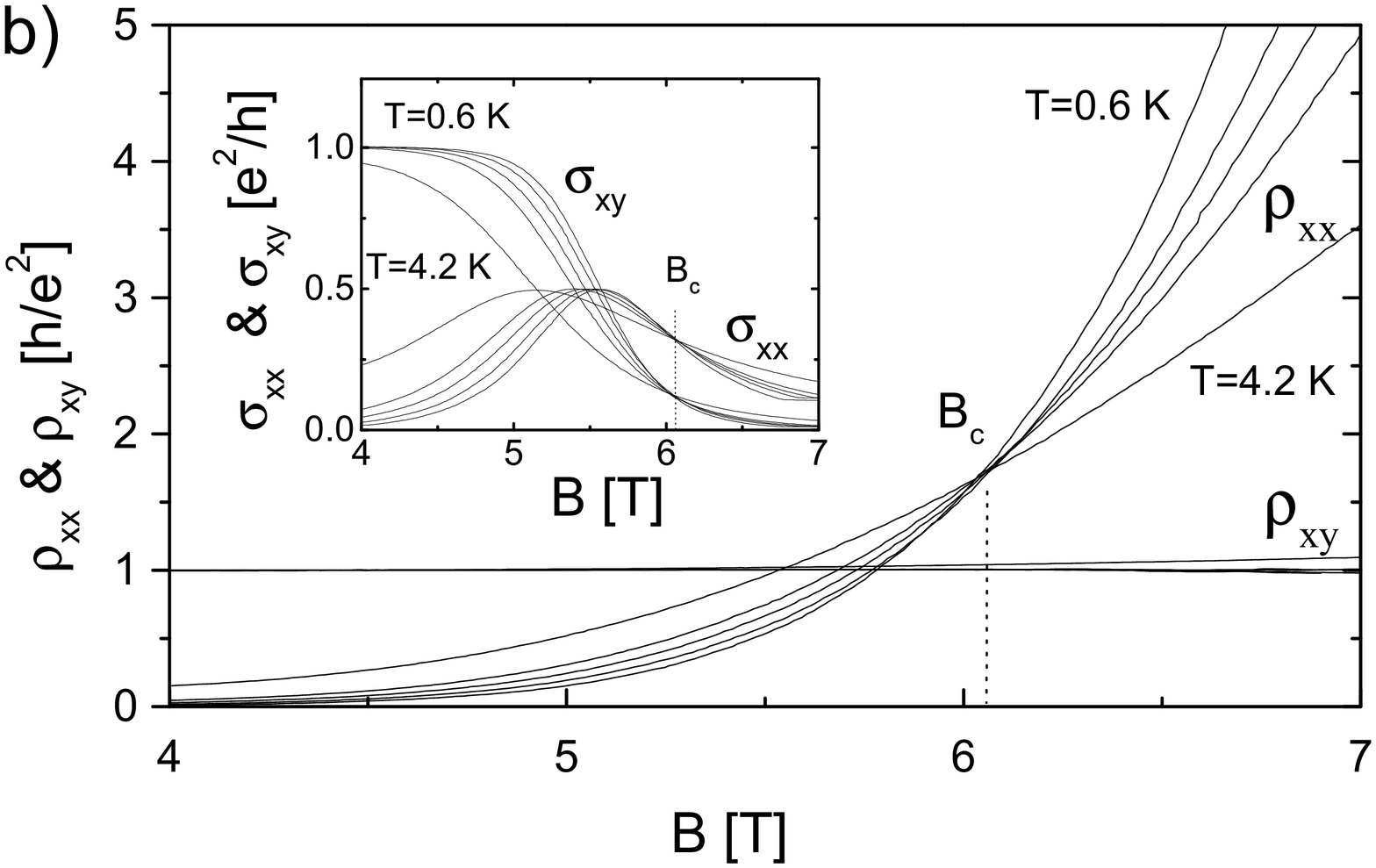} \vspace*{-1cm} \caption{Magnetic field
dependence of $\rho_{xx}$ and $\rho_{xy}$ for temperatures ranging
between 0.6 K and 4.2 K. In the insets, the conductivities
obtained from $\rho_{xx}$ and $\rho_{xy}$ are plotted as a
function of the magnetic field. $B_C$ and $B_C^L$ are the critical
magnetic fields for which $\rho_{xx}$ is temperature independent
($B_C^L=1.3 T$ and $B_C=6.1 T$). Fig. a) and b) were obtained from
different cool-downs and densities ($n=1.2\times 10^{11}cm^{-2}$
and $n=0.8\times 10^{11}cm^{-2}$.}
\end{figure}

When extracting the transition from the crossing point we imply
that only two types of states exist in our system: a quantum Hall
and an insulating state. The crossing point, which defines a
$T$-independent point, can be reliably taken as the transition
point. However, in recent experiments a zero-field metallic-like
state in a 2D system was observed \cite{krav94}. In these systems,
for high enough densities and at $B=0$ the resistance decreases
and saturates with decreasing $T$, which invalidates the use of
the crossing point to extract the boundary. Therefore,
alternatively the peak in the conductivity was used to determine
the transition \cite{krav,dultz}. The question of the existence of
only an insulating state at $B=0$ is even more fundamental. In our
Ge/SiGe system the insulating behavior is well defined in the low
density limit with an increase of $\rho_{xx}$ by more than an
order of magnitude when $T$ is lowered from 1K to 0.3 K. This is
in contrast to our highest density, where the increase in
$\rho_{xx}$ is only a few percent for the same $T$ span. Are these
two behaviors characteristics of the same insulator? If yes, our
procedure is well defined. On the contrary, if these two regions
belong to different zero-$T$ phases one could expect signatures at
our experimental $T$. However, we do not observe any qualitative
change of the $T$-dependence in our entire density range, which
indicates a transition or cross-over to a metallic-like state at
$B=0$. This is correlated with the existence of a well defined
crossing point as a function of $B$. On the quantum Hall side we
have the reverse $T$-dependence with a similar magnitude close to
the transition.

The next step is to obtain the entire phase diagram, by varying
the density and $B$. In most previous experimental phase diagrams
the degeneracy of the Landau levels could not be resolved down to
the insulating phase. Typically the Zeeman splitting was much
smaller than the cyclotron gap \cite{wong}. In holes there is an
additional degeneracy of the valence band, which leads to a two
band system, which affects the phase diagram \cite{krav,dultz}. We
avoid these difficulties because our 2DHS is confined in a
strained Ge layer, where the strain removes the degeneracy of the
valence band. The 150 \AA \ thick Ge layer is sandwiched in
between Si$_{0.4}$Ge$_{0.6}$ layers, where Boron modulation doping
is placed. The crucial property of this system is that the Zeeman
activation energy is only 30 \% smaller than the cyclotron
activation energy. As a result, the Zeeman energy is resolved down
to our lowest density so that we observe all even and odd filling
factors. To vary the density we use a top gate with an insulating
silicon oxynitride layer grown between the cap layer and the gate.
By applying a gate voltage between 0 V and 6 V we could vary the
density between $n=0.5-5\times 10^{11}\mbox{cm}^{-2}$ with no
leakage. The highest mobility we obtained was $20\times 10^{3}
\mbox{ cm}^2/Vs$ at zero gate voltage. The measurements were
performed in a dilution refrigerator and a He$_3$ system at {\em
T} 's ranging between 40 mK and 4.2 K, using standard AC lock-in
technique with an excitation current of 0.1 nA. DC measurements
were performed in order to check for consistency.

We determine the transition between two quantum Hall liquids from
the local Max\{$ d\rho_{xy}/d B$\} \cite{hp} at 40 mK. This point
is {\em T}-independent at low $T$ and the size of the dots are
representative of the differences obtained from the 200mK and
400mK curves. However, other choices for determining the
transition are possible, such as Max\{$\rho_{xx}$\} or Max\{$
d\sigma_{xy}/d B$\} \cite{shahar4}. When using these other methods
we found that they give similar results. The insulator-to-quantum
Hall transitions (solid dots) were extracted from the crossing
points of $\rho_{xx}$ at three different {\em T} 's (40mK, 200mK
and 400mK), which also gives an estimate of the experimental
inaccuracy represented by the size of the dots. The results are
summarized in fig. 2.

\input epsf
\begin{figure}
\epsfysize=9cm \epsfbox{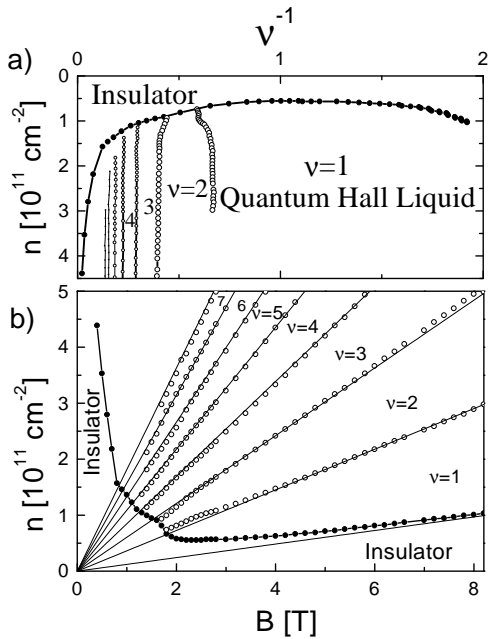} \caption{a) the phase diagram in
terms of the inverse of the filling factor. Solid dots are
insulator-to-quantum Hall transitions and open dots quantum
Hall-to-quantum Hall transitions. b) the Landau level fan.}
\end{figure}

At a fixed $\nu$ and with decreasing density, there are direct
transitions between the quantum Hall phases $\nu=1,2,3,...$, and
the insulating phase. This is the main result of fig. 2a and is in
strong contrast to KLZ's GPD, where the insulating phase is only
bordered by the $\nu=1$ quantum Hall liquid. These results are
very similar to those obtained in Si MOSFET \cite{krav}, where a
merging of the transition states into the insulating phase and an
absence of floating was observed. However, in Si MOSFET each
Landau level is doubly degenerate due to the additional valley
band degeneracy. Therefore, the $\nu=5$ state corresponds to the
second Landau level. Other results with Si MOSFET (using an
arbitrary cut off value of the conductivity to determine the
transition) were interpreted as consistent with the floating up
scenario \cite{shashkin,okamoto} like in n-type GaAs/AlGaAs
\cite{glozman}. p-type GaAs/AlGaAs has a similar valley band
degeneracy, but no floating was observed and mainly even filling
factors were resolved \cite{dultz}. An interesting result was
obtained by Lee {\em et al.}\cite{chang}, who used a multiple GaAs
quantum well and obtained similar transitions for high even
filling factors. As an alternative method to determine the
transition, the disappearance of the activated behavior was used
in ref. \cite{pudalov}. Transitions up to $\nu=3$ were observed in
a similar Ge/SiGe system by Song {\em et al.}\cite{song}.

To analyze the termination of the Landau levels as a function of
{\em B}, we plotted in fig. 2b the same data as in fig. 2a, but as
a function of {\em B}. At high $B$, the experimental points (open
and solid dots) follow very precisely the
$n=(i+\frac{1}{2})\frac{eB}{h}$ lines, where $i=0,1,2,...$. The
insulating phase below the lowest energy level ($i=0$) can be
understood in terms of simple energy level physics. The energy
levels are broadened by disorder, which leads to a smooth density
of states around them and to the formation of mini-bands. The
states at the center of the mini-bands are extended, but the
states with energies away from the band center are localized
\cite{janssen}. In this way, at $T=0$ and when the Fermi energy
($F_F$) is below the lowest energy level, $E_F$ is pinned to the
localized states and the system becomes insulating. The deviation
from the linear dependence ($i=0$), when $B$ is decreased, is due
to the disorder broadening. Lowering $B$ decreases the gaps and
leads to an overlap of the energy levels. Therefore, some states
from higher energy levels increase the density of states below the
lowest energy level ($i=0$), which leads to the apparent
``floating up'' of the lowest energy level, when $E_F$ is pinned
to the $i=0$ level. When assuming a symmetric broadening, the
transition occurs at $\nu=1/2+\nu_h$. In this case, $\nu_h$ is
simply the density of states of the higher energy levels
(normalized to one energy level) integrated between zero and the
$i=0$ level. Hence, $\nu_h$ is a measure of the overlap between
the density of states of higher energy levels and the lowest
energy level. Applying the same argument to higher energy levels
does not hold, because the contribution of states from even higher
levels is compensated by the states lost from lower levels.
Therefore, broadening cannot affect the higher energy levels and,
indeed, we observe {\bf no floating} of higher energy levels.
These higher energy levels simply merge straight into the
insulating phase.

Before discussing our results in relation to existing theoretical
ones, we need to make an important point. Most theories, starting
with KLZ's GPD, are a function of the {\em B}-field strength and
the disorder strength. At first sight our system is very
different, as we replaced the disorder axis by the density. One
could argue that, when the density is reduced, screening becomes
smaller. However, in two dimensions and at $B=0$ the screening is
largely density independent. The situation is slightly different
with a quantizing {\em B}-field because the screening becomes
non-linear \cite{efros}. In addition, screening is also
$T$-dependent. On the other hand, it is the ratio of the disorder
potential fluctuation over the quantizing energy, which can be
effectively tuned by the density. Therefore, our results suggest
that at strong level mixing (which can be estimated by $\nu_h$),
the extended states at the center of each energy level disappear
and localization takes over. This observation is in agreement with
recent theoretical and numerical calculations. Fogler, for
instance, showed that the levitation of extended states remains
very weak even at low {\em B} \cite{fogler}. Using tight binding
models, Liu {\em et al.} \cite{liu} and more recently Sheng and
Weng \cite{sheng,sheng3} showed evidence for the disappearance of
extended states without floating. Inspired by recent experimental
results, a numerical phase diagram based on the tight binding
model has been obtained \cite{sheng2}. The similarity between
their GPD and ours (fig. 2) is very striking.

\input epsf
\begin{figure}
\epsfysize=6cm \epsfbox{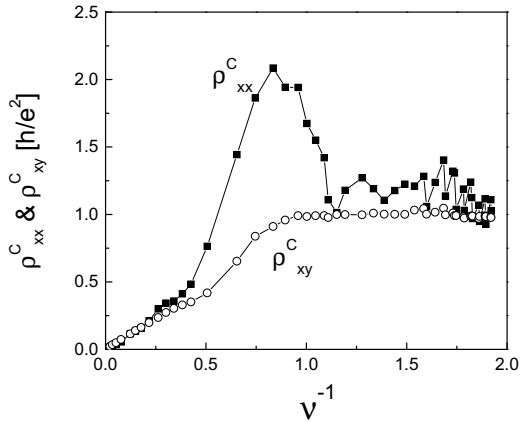} \caption{Inverse filling factor
dependence of $\rho_{xx}^C$ and $\rho_{xy}^C$.}
\end{figure}

We now turn to the analysis of the resistivities at the
transition. In fig. 3, we have plotted the critical resistivities,
$\rho_{xx}^C$ and $\rho_{xy}^C$ defined as the value of the
resistivity at the quantum Hall liquid-to-insulator transition.
This work was inspired by a recent letter by Song {\em et al.}
\cite{song}, who obtained a similar curve for $\nu>1$. Our
striking new result is the peak of $\rho_{xx}^C$ close to $\nu=1$
before $\rho_{xx}^C$ saturates around the quantized value $h/e^2$.
In addition, $\rho_{xy}^C$ follows the classical expression for
$\nu>1$ indicative of a Hall insulator, which becomes closely
quantized for $\nu<1$, bordering the quantized Hall insulator. A
similar peak in $\rho_{xx}^C$ was obtained very recently by Hanein
{\em et al.} \cite{hanein} in systems exhibiting a zero {\em B}
metal-insulator transition. But in their case $\rho_{xx}^C$ tends
to $h/e^2$, with vanishing $B$, as opposed to zero in our case.
The physical origin of this peak is not understood.

Summarizing, we have experimentally mapped out the phase diagram
of the integer quantum Hall effect, as a function of density and
magnetic field. At low fields and close to the quantum Hall
liquid-to-insulator transition, we have observed the floating up
of the lowest energy level, but no floating of any higher levels,
rather a merging of these levels into the insulating state. These
results are consistent with the disappearance of extended states
without levitation due to the broadening of the energy levels.
Along the transition, we observe a peak in the critical
resistivity around filling factor one.

We would like to acknowledge M.M. Fogler, H.W. Jiang, S. Kivelson
and D.N. Sheng for helpful discussions. This work was supported in
part by the National Science Foundation.

\end{document}